\documentstyle[twocolumn,aps,prb]{revtex}
\begin{document}
\draft

\title{Interlayer Exchange Coupling in Fe/Cr multilayers }

\author{L. Tsetseris, Byungchan Lee, and  Yia-Chung Chang}
\address{Department of Physics and 
Materials Research Laboratory 
\\ University of Illinois at 
Urbana-Champaign, Urbana, Illinois 61801}

\date{\today}
\maketitle


\begin{abstract}
 We investigate the origin of  the long period 
 oscillation of the interlayer exchange coupling 
 in Fe/Cr trilayer systems. Within the stationary phase
 approximation the periods of the oscillations are
 associated with extremal vectors of the Fermi
  sphere of Cr. Using a realistic tight-binding model with spin-orbit
interaction we calculate the coupling strength for each extremal vector 
  based on the spin-asymmetry of the reflection amplitude for a propagating
state impinging from the Cr to Fe layer.
 We find that for (001) and (110) growth 
 direction the biggest coupling strength comes from the 
 extremal vector centered at the ellipsoid N of the 
 Fermi surface of Cr.

\end{abstract}

\section{Introduction}
 The magnetic properties of multilayer systems have 
 been a subject of intensive study, particularly  for 
 the last decade. The initial observation that two 
 ferromagnetic layers can be coupled 
 antiferromagnetically, when separated by certain type 
 of magnetic or non-magnetic spacer 
  (Gr\"{u}nberg $et \ al.$~\cite{Grunberg}),  
 and the following discovery that this coupling possesses 
 a damping oscillatory behavior (Parkin $et \ al.$~\cite{Parkin1})  
 stimulated a great deal of interest~\cite{Hein}. 
 Moreover, because of the  
 phenomenon of giant magnetoresistance (GMR)~\cite{gmr1,gmr2},  
 these structures have become very 
 promising candidates for applications in the
 magnetic recording industry.

 The construction of ultrathin structures poses great 
 challenges from the experimental point of view. The initial 
 samples grown with sputtering displayed the characteristic 
 oscillatory behavior, however, soon after the first 
 measurements, the demand for strict control over stoichiometry 
 and disorder made the growth of excellent epitaxial 
 sandwiches or superlattices necessary. In the case of a 
 sandwich, it is a common practice now that
 one uses a wedge geometry for the spacer layer~\cite{Unguris}. In this 
 way one can study the spacer thickness dependence of the 
 interlayer exchange coupling (IEC) using only one sample.
 
 From the theoretical perspective, the calculation of IEC has also 
 attracted a lot of attention. Several different approaches 
 to the problem have been suggested and used extensively. 
 One can divide all these approaches into two major classes. One 
 class consists of total energy calculations, in which
 one calculates the difference in the total energy of the 
 sample for the two important magnetization configurations 
 (the magnetization of the ferromagnetic layers being parallel 
 in one case, antiparallel in the other). Local spin density 
 approximation (LSDA) (see for example Refs.~\onlinecite{Herman,Schilf})
and semiempirical tight-binding (TB)
 methods~\cite{Hase,Stoef}  have been used for this purpose.
 The second class consists of various model calculations. 
 In this we include the calculations based on the 
 perturbative treatment of the RKKY interaction~\cite{Yafet,Bruno1},
 an adapted Anderson (or sd-mixing) model~\cite{sd-mix1,sd-mix2}, 
 and perhaps most importantly, we 
 include all the various calculations of IEC that exploit 
 the idea of quantum interference and quantum confinement in 
 the spacer material~\cite{Edwards,Bruno2}. The multiple reflections that 
 the electrons experience at the spacer/magnetic layer 
 interfaces bear an analogy to Fabry-Perot like interferometry.
 Our own method belongs to this class. All the model 
 calculations have certain approximations built in, and hence,  
 one has to be aware that in some cases the accuracy of
 the results is limited. Nevertheless, the potential for 
 numerical errors involved in these calculations is 
 significantly smaller than in the total energy 
 calculations, particularly for large spacer thickness, and   
 the computational effort is much less as well.
 
 It is now widely accepted that 
 the periods of the oscillations of IEC are given by the
 extremal vectors of the Fermi surface of the spacer material. 
 In this paper we are going to study IEC of Fe/Cr/Fe, so 
 the spacer is Cr. Chromium has a complicated Fermi surface with 
 a rich variety of spanning features.  
 Therefore one would expect in principle a multiperiodic
 oscillation of IEC with respect to the Cr layer thickness. 
 However, only two periods have been observed in the experiments 
 so far. The short period is roughly equal to two monolayers 
 (ML) of chromium and it is believed to be due to the
 nesting vector that gives rise to antiferromagnetism in Cr. 
 The origin of the long period (measured to be somewhere between 
 15~\AA \ and 18~\AA  ) has been proven a more difficult problem. 
 
 We use an empirical TB method in order to calculate the coupling 
 strengths associated with different periods. By using the 
 ``force theorem''~\cite{Mack}, one can find the change 
 in the density of states for the trilayer system with respect 
 to the bulk case. Electrons with different spin see 
 different effective potentials at the Fe/Cr interface, and 
 consequently, the multiple reflections are sensitive to the
 magnetizations of the Fe layers being parallel or antiparallel.
 The base of the analytical work was developed in a series of 
 papers by Bruno~\cite{Bruno2}. Stiles~\cite{Stiles1} 
 and Slonczewski~\cite{Slonc} also showed how one can calculate 
 IEC based on the spin asymmetry of the reflection amplitudes. 
 In more recent papers the method was applied with
 encouraging results in real systems~\cite{BrSt}. We have used the 
 same method as Lee and Chang used for Co/Cu/Co systems~\cite{Lee}, but with 
 some modifications, in order to take care of the much more 
 complicated Fermi surface of Cr. The main conclusion of our work 
 is that the long period oscillation originates from the 
 extremal vector at the ellipsoid centered at point N of the Cr 
 Fermi surface. This is true for both (001) and (110) growth 
 directions. The result agrees with that obtained with LDA 
 calculations by Stiles~\cite{Stiles2}. We will also discuss its validity 
 in view of the outcome of recent photoemission experiments~\cite{Li}.

\section{Theoretical Model}

 The Interlayer Exchange Coupling is defined as
\begin{equation}
 J=\frac {\Omega_{F}-\Omega_{AF}} {2 S},
\end{equation}
 where $\Omega_{F}$ and $\Omega_{AF}$ are the grand 
 canonical potentials for the ferromagnetic (F) and 
 antiferromagnetic (AF) configurations and $S$ is the
 area of the sample. The electrons are partially reflected at
 the interfaces of the Fe/Cr/Fe sandwich,
 and the change in the grand  canonical 
 potential is given by the force theorem (or the frozen 
 potential approximation) as
\begin{equation}
\Delta \Omega_{\nu} = \frac {1}{\pi} {\rm Im} \sum_{{\bf k}_{\|}} 
\int_{-\infty}^{+\infty} 
d\epsilon \,f(\epsilon)\, Tr\, \ln(1-G_{0} T^{L,\nu} 
G_{0} T^{R,\nu}),
\end{equation} 
 where $\nu$ labels the two configurations (F and AF), 
 $G_{0}$ is the bulk Green's function of the spacer 
 material, $T$ is the $T$-matrix, $f$ is the Fermi-Dirac 
 distribution , and $L$ and $R$ stand for 
 the left and right interface respectively. $Tr$ \ denotes the 
 trace over the z-component of the wavevector and the
 spin index.
 The asymptotic form of IEC (for large thickness D of 
 the Cr layer) is given in first order by the simple 
 analytical result (see Ref.~\onlinecite{Lee} for details)
\begin{equation} 
J= \,\, {\rm Im} \sum_{\alpha} \sum_{ij} \frac {\hbar u_{ij}^{\alpha} 
\kappa_{ij}^{\alpha}}
{4\pi^{2} D^{2}} \, \Delta R_{ij}^{\alpha} \,
 e^{i(q_{ij}^{\alpha} D + \phi_{ij}^{\alpha})} 
 \, F_{ij}^{\alpha}(D,T) \, \theta_{ij},
\end{equation}
 where 
\begin{equation}
 \Delta R_{ij}^{\alpha}= \sum_{\sigma \sigma'}
 (r_{R\,ij}^{\sigma \sigma' F} 
 r_{L\,ji}^{\sigma' \sigma F}-
 r_{R\,ij}^{\sigma \sigma' F} r_{L\,ji}^{\sigma' \sigma AF}),
\end{equation}
\begin{equation} 
F_{ij}^{\alpha}(D,T)=\frac {2\pi k_{B} T D/\hbar 
u_{ij}^{\alpha}} {\sinh(2\pi 
k_{B} T D/\hbar u_{ij}^{\alpha})},
\end{equation}
and 
\begin{equation}
\frac{1}{u_{ij}^{\alpha}}=\frac {1}{u_{i}^{\alpha}} - \frac {1}
{u_{j}^{\alpha}}.
\end{equation}
 In Eq.~(3), $q_{ij}^{\alpha}=k_{j}^{\alpha}-k_{i}^{\alpha}$ 
 is an extremal vector of the Fermi surface of Cr
 parallel to the growth direction (which is taken 
 to be the $\hat{z}$ direction). The superscript $\alpha$ 
 labels different ${\bf k}_{\|}^{\alpha}$, that is the in-plane component
 of the wavevector. $k_{i}^{\alpha}$ 
 and $k_{j}^{\alpha}$ are the z-components of the wavevector 
 of the incident (reflected) and reflected (incident) electron 
 for the right (left) interface, and  $u_{i}^{\alpha}$ 
 is the group velocity at a point $({\bf k}_{\|}^{\alpha},
 k_{i}^{\alpha})$ of the Fermi 
 sphere. The index $\sigma$ labels the two spin mixed states
 that are degenerate at each $({\bf k}_{\|}^{\alpha},k_{i}^{\alpha})$
 (Kramers degeneracy).
 The reflection amplitude from a state $(k_{i}^{\alpha},\sigma)$
 to a state $(k_{j}^{\alpha},\sigma')$ is denoted by 
 $r_{I\,ij}^{\sigma \sigma'\, \nu}$ (where $I$\,=\,$L$,\,$R$ labels
 the interfaces). 
 $\kappa_{ij}^{\alpha}$ is related to the curvature 
 radii at the two endpoints of $ q_{ij}^{\alpha}$, 
\begin{equation}
\kappa_{ij}^{\alpha}= \left[\sqrt{\left| \frac{\partial^{2} q_{ij}^{\alpha}}
 {\partial k_{x}^{2}} \frac {\partial^{2} q_{ij}^{\alpha}}
 {\partial k_{y}^{2}} - \left(\frac {\partial^{2} q_{ij}^{\alpha}}
 {\partial k_{x} \partial k_{y}}\right)^{2} \right| } \,  \right]^{-1} 
 \end{equation}
 The product 
\begin{equation}
G_{ij}^{\alpha}=\frac {\hbar} {2\pi^{2}} u_{ij}^{\alpha}
  \kappa_{ij}^{\alpha}
 \end{equation}  
 is called the geometrical weight of the extremal vector 
 $ q_{ij}^{\alpha}$. 
 D is the thickness of the Cr layer, $\theta_{ij}$ is 
 a constraint function, which 
 is 1 for extremal vectors and 0 otherwise. The phase factor 
 $\phi_{ij}^{\alpha}= 0$,
  $\pi/2$, and $\pi$ for maximum, saddle, and minimum 
 extremal points respectively. We perform the calculation 
 at zero temperature,  
 so $ F_{ij}^{\alpha}(D,T)=1$. For finite temperatures
 $ F_{ij}^{\alpha}(D,T)$ is less than 1, if the
 spacer is magnetic~\cite{Bruno2}.
  
 The reflection amplitude factor $\Delta R_{ij}^{\alpha}$  
 depends explicitly on the reflection 
 amplitudes, which, as we mentioned above, are calculated 
 from an empirical TB model with $s$, $p^{3}$, 
 and $d^{5}$ orbitals. Because of the sandwich geometry one  
 needs to include, not only the propagating Bloch solutions, but 
 also the evanescent states, i.e. the states with complex 
 wavevectors. The details of the method can be found in Ref.~\onlinecite{Chang}.
 For Cr, the spin-orbit coupling is significant only for the 
 so called lens area 
 and at the point across the $\Gamma$H line where the electron
 surface almost touches the hole surface (see Fig.~\ref{f:fig1}). For all
 other vectors the spin-orbit interaction does not play an
 important role, the spin is
 approximately a good quantum number, and one finds that
\begin{equation} 
 \Delta R_{ij}^{\alpha}\approx (r_{Rij}^{\alpha +}-r_{Rij}^{\alpha -}) 
 (r_{Lji}^{\alpha +}-r_{Lji}^{\alpha -}) ,
 \end{equation}
 where the superscript + ($-$) denotes majority (minority)
 spin electron.

\noindent \section{ F\lowercase{e}/C\lowercase{r}/F\lowercase{e} sandwiches} 
 
 As we pointed out in Sec.~I, the Fe/Cr/Fe case has 
 been a difficult one for the calculation of IEC, and 
 this is mainly because of the complicated Fermi surface of Cr. 
 For our own TB calculations we have used 
 the parameters from Ref.~\onlinecite{Papa}. These parameters do not 
 include any relativistic effects, and particularly spin-orbit 
 coupling. We have included spin-orbit coupling with its 
 characteristic parameter $\xi=2.84$ mRy taken from Ref.~\onlinecite{Herman2}. 
 This value agrees well also with the value found in Ref.~\onlinecite{Mack}. 
 The (100) cross section of the
 Fermi surface, that we obtained, is shown in Fig.~\ref{f:fig1}. 
 In this figure we used
 a slightly larger parameter $\xi$, so that the effects of the spin orbit
 interaction become easier to see. 
 The Fermi surface we obtained (with the correct $\xi$)
 agrees very well qualitatively, and in most cases 
 quantitatively, with the ones obtained from experiments and 
 other more involved band structure calculations (see for example 
 Ref.~\onlinecite{Fawc}). 
 The inclusion of spin-orbit coupling affects significantly 
 the area around the lens vector, as we mentioned before. One needs to 
 take this effect into consideration, because the period 
 related to this vector is very close to the long period 
 of the oscillations of IEC.

 Our results are summarized in Tables~\ref{table1} and~\ref{table2} for the two 
 growth directions that we have studied. In the tables, 
 $c=\frac{a}{2},\frac{a}{\sqrt{2}}$ 
 is the distance between two Cr layers for (001) and (110) 
 orientation respectively ($a$\,=\,2.88~\AA \ is the Cr lattice constant), 
 and $\lambda$\,'s are the periods that we found.  
 In the last column we give 
 the coupling strength 
\begin{equation} 
 J_{ij}^{\alpha}=\frac {G_{ij}^{\alpha} n |\Delta R|}
  {2 D^{2}}
 \end{equation}
 associated with each vector $q_{ij}^{\alpha}$. $G_{ij}^{\alpha}$ is the 
 geometrical weight and $n$ is the number of equivalent 
 extremal vectors
 in the first Brillouin Zone. For the thickness 
 D we take a typical value of 10~\AA. One can see that 
 $|\Delta R|$ and $G_{ij}^{\alpha}$ vary significantly for 
 different extremal vectors, and therefore, the overall coupling strength 
 is not negligible only for very few vectors. In the next two 
 subsections we shall discuss the analyses for (001) and (110) 
 orientations separately.

\noindent \subsection{(001) orientation}

 In Fig.~\ref{f:fig1}, we give the (100) cross section of the Cr Fermi 
 sphere. The wavevectors are measured in $\frac {2 \pi} {a}$. 
 We have marked some extremal vectors, 
 which, according to the theory, are candidates for the observed 
 periods of the oscillations of IEC. Four vectors are of
 particular importance. One is the vector ($\overline{N_{1}}$)
 at the ellipsoid pocket at N. Two other special vectors are the 
 one spanning the lens ($\overline{L_{1}}$) and the one just 
 outside the lens ($\overline{L_{2}}$). Finally, we have  
 the nesting vector ($\overline{V_{n}}$), 
 which connects two almost parallel lines, one from 
 the electron octahedron around point $\Gamma$, and the other from 
 the hole octahedron around point H. 

 For the extremal vector $\overline{N_{1}}$, 
 both $G_{ij}^{\alpha}$ and $|\Delta R|$ are big, and hence,  
 the coupling strength is large (7.94~mJ/m$^2$). 
 The next closest amplitude is 2.80~mJ/m$^2$, for the extremal 
 vector $\overline{X}$ that spans the electron 
 ball at point X [${\bf k}_{\|}=(0,0.42) 
 \frac {2 \pi} {a}$]. This is still 2.83 times smaller than
 the ellipsoid contribution. All the other amplitudes, with the exception
 of the nesting vector, are, approximately, at least 
 one order of magnitude smaller.  
 In Figs.~\ref{f:fig2}(a)-(b), we present the behavior of the modulus of the
 reflection amplitude \,$|r^{+}|$ \, ($|r^{-}|$) of the majority 
 (minority) spin electron as a function of the energy
 for ${\bf k}_{\|}=(0,0.5) \frac {2 \pi} {a}$ and
 ${\bf k}_{\|}=(0,0.42) \frac {2 \pi} {a}$. For these ${\bf k}_{\|}$'s, 
 the spin-orbit coupling is not very important and $|\Delta R|$ 
 is given explicitly by the square of 
 the difference of $r^{+}$ and $r^{-}$. These two 
 complex numbers mirror
 the match or the mismatch of the bands of Cr and Fe at the Fermi 
 level and at this particular ${\bf k}_{\|}$. The band structure along 
 $k_{z}$ at ${\bf k}_{\|}=(0,0.5) \frac {2 \pi} {a}$ for Cr and Fe is given 
 in Fig.~\ref{f:fig3}. For both Fe and Cr the Fermi level is aligned to be
 at zero energy.
 
 One can see from Fig.~\ref{f:fig2}(b) [${\bf k}_{\|}=(0,0.42) 
\frac {2 \pi} {a}$] 
 that $|r^{+}|$ changes very
 rapidly around the Fermi energy. This means that the stationary
 phase approximation is not very well satisfied for this ${\bf k}_{\|}$. 
 Therefore, the result for the coupling strength for $\overline{X}$
 is not very reliable.
 For $\overline{N_{1}}$ on the other hand, 
 the variation of $|r^{+}|$ and $|r^{-}|$ around the Fermi level is smooth.
 The majority electrons of the ellipsoid
 are reflected strongly at the interface, because
 they have to transmit from a Cr $sp$-band to an Fe $d$-band 
 (see Fig.~\ref{f:fig3}). 
 The minority electrons have smaller reflection amplitude
 $|r^{-}|$ because the related Fe band at the Fermi level is of the same
 character as the Cr band. As soon as the energy is raised by
 $\approx 0.01$ Ry, the Fermi level minority band of Fe changes symmetry,
 and the minority electrons become completely confined at the Cr layer. 

 Our calculations give a period of 13.77~\AA \ for $\overline {N_{1}}$.  
 This period is slightly smaller than the experimental 
 value for the long-period oscillation, which is somewhere 
 between 14~\AA \ and 18~\AA \ (see for example 
 Refs.~\onlinecite{Parkin1,Unguris,Purcell}).
 For this same vector, the
 de Haas--van Alphen experiments~\cite{Graebner} 
 give a period of 15.97~\AA,  
 which is within the experimental uncertainty. With a satisfactory 
 period and coupling strength (as we will argue in the Sec.~IV), 
 the vector $\overline {N_{1}}$ is the best candidate for the origin 
 of the long period oscillations of IEC.

 The very existence of the lens is due to the spin-orbit 
 coupling that we have included for Cr. In the case of vanishing 
 spin-orbit parameter $\xi$ the lens touches the outside surface 
 and there is no extremal vector at this point. With $\xi= 2.84$ mRy 
 we obtained two extremal vectors $\overline{L_{1}}$
  and $\overline{L_{2}}$ at ${\bf k}_{\|}= 
 (0,0.29) \frac {2 \pi} {a}$, which have periods 20.85~\AA \ and 17.77~\AA 
 \ respectively.  For both vectors, the factor $|\Delta R|$ 
 is not small, but in contrast to the nesting vector case, the 
 curvature radius factor $\kappa_{ij}^{\alpha}$ is very small. 
 Thus the overall contribution to the coupling is almost negligible. 
 This is a somehow unwelcome result because the periods of 
 $\overline{L_{1}}$ and $\overline{L_{2}}$ 
 are very close to the experimental one. However, it 
 is a result that one would anticipate from qualitative reasoning 
 alone, since the phase space associated with the lens is small, 
 as has been stressed by other authors previously~\cite{Schilf}.

 The nesting vector $\overline{V_{n}}$ corresponds to a period
 of about 3.12~\AA. This is about 2.17~ML of Cr 
 and agrees very well with previous
 results. For this vector, the geometrical weight is very big 
 (approximately at least an order of magnitude bigger than 
 most of the other $G_{ij}^{\alpha}$'s presented in Table~\ref{table1}). This 
 is as expected qualitatively, because the nesting feature (nearly parallel 
 lines) leads to very large value for $\kappa_{ij}^{\alpha}$. 
 Since the reflection factor $|\Delta R|$ is not very small, 
 the overall coupling strength associated with $\overline{V_{n}}$ 
 is large  (37.99~mJ/m$^2$). Although this number is too big compared
 to experimental strengths, the result agrees 
 with the experiment qualitatively, since it renders $\overline{V_{n}}$  
 the best candidate for 
 the short period oscillations (for which the period is  
 measured to be 2.1~ML~\cite{Unguris,Purcell}).
 Nevertheless, one has to acknowledge that the nesting feature
 requires a treatment that goes beyond  the stationary phase 
 approximation used in our calculations. Specifically the 
 replacement of the Fermi surface sheets by parabolas is  
 questionable for this case, and one should perform an explicit numerical 
 integration over ${\bf k}_{\|}$ for this particular area of the Fermi 
 sphere. Moreover, the coupling strength of the short period 
 oscillations is very sensitive to roughness, as will be 
 discussed in Sec.~IV. 

\noindent  \subsection{(110) orientation}

 In Table~\ref{table2} we give the results of
 the calculation for the (110) orientation. 
 Again, as in the (001) case, one can find, among others, extremal
 vectors that span the lens ($\overline 
 {L_{1}'}$, $\overline {L_{2}'}$, and $\overline {L_{3}'}$) and vectors 
 (for example $\overline {N_{1}'}$) at the N ellipsoid.  
 As we can see in this case, the most important 
 contribution to the coupling is made by $\overline {N_{1}'}$. 
 Its period and strength are found to be 16.46~\AA \ (the experimental
 period is $\approx$ 18~\AA~\cite{Parkin1,Parkin2}) and
 4.65~mJ/m$^2$ respectively. 
\nopagebreak
 In Fig.~\ref{f:fig4}, we give the moduli of the reflection 
 amplitudes $|r^{+}|$ and $|r^{-}|$ as a function of energy. 
 The majority electrons
 are strongly confined in the spacer, whereas the minority electrons are
 only partially reflected at the interface. The change of $|r^{+}|$ and 
  $|r^{-}|$ around the Fermi level is slow enough, that the stationary 
 phase approximation is reliable. 

 The lens vectors $\overline {L_{1}'}$, $\overline {L_{2}'}$, 
 and $\overline {L_{3}'}$
 have again small $\kappa_{ij}^{\alpha}$, and small overall coupling 
 strengths.  A significant contribution is made from the vector 
 $\overline {X'}$ at ${\bf k}_{\|}=(0,0.44) \frac {2 \pi} {a}$. Note 
 that we have used new $x',\, y',\,z'$ \ axes so that $\hat{z}'$ is the growth 
 direction. 
 Another ellipsoid vector at ${\bf k}_{\|}=(0.35,0.50) \frac {2 \pi} {a}$ 
 has also large $J_{ij}^{\alpha}$. Anyway, these
 two $J_{ij}^{\alpha}$'s (0.89~mJ/m$^2$ \ and 1.17~mJ/m$^2$) are 5 and 4  
 times smaller than that of $\overline {N_{1}'}$. Therefore, the origin 
 of the long period oscillations is attributed again 
 to an ellipsoid spanning vector.
 There is also a non-negligible strength associated
 with short period oscillations. This is for ${\bf k}_{\|}=(0.0,1.0) 
 \frac {2 \pi} {a}$, with a period of 4.70~\AA \ and a strength
 of 2.02~mJ/m$^2$. To the best of our knowledge, no short period
 has been reported so far for the (110) orientation.

\section{Discussion of results}
 
 As we stressed in the previous sections, the two most important issues
 that one has to address for IEC are the period of the oscillation
 and its magnitude. For the three cases of (001), (110), and 
 (211) orientation, the experimental data give a similar period. 
 The common period is in 
 favor of the 
 argument that the large wavelength oscillations are originated from
 a relatively isotropic part of the Cr Fermi surface. As we saw above, 
 our calculations agree with that, since the two vectors  
 $\overline {N_{1}}$ and $\overline {N_{1}'}$ are extremal vectors of the
 ellipsoid centered at N, which is fairly isotropic. The size of the
 ellipsoid, as we calculated it in the (110) case (period 16.46~\AA) or
 as it can be calculated from de Haas--van Alphen experiments for
 the (001) case (period 15.97~\AA), 
 is in good agreement with the observed period.
 
 The measured coupling strengths for the (001) orientation 
 vary from 0.6~mJ/m$^2$ to 1.6~mJ/m$^2$ for thicknesses 
 between 4 and 8~ML~\cite{Purcell,Fullerton,Fert}.
 Our own result is 7.94~mJ/m$^2$ for D\,=\, 10~\AA, if we use Eq.~(10),
 whereas if we use Eq.~(3), we find that the first antiferromagnetic peak 
 strength is 14.45~mJ/m$^2$
 for D\,=6.7~\AA. These numbers are about 10 times
 the experimental ones. Previous total energy calculations (see section
 2.1 in Ref.~\onlinecite{Hein}) 
 predicted larger amplitudes (up to 100 bigger than the experiment),
 so our result is an improvement. In a most recent 
 first principle calculation, Stiles~\cite{Stiles2} also attributed
 the origin of the long period oscillations to 
 $\overline {N_{1}}$ and $\overline {N_{1}'}$. His 
 calculated coupling strengths for these vectors are
 5.7~mJ/m$^2$ and 3.2~mJ/m$^2$. Both numbers are in good agreement with ours.
 
 The discrepancy with the experiment can be remedied if the interface roughness
 is taken into account. The coupling strength in realistic samples decreases 
 mostly because of the interface roughness. 
 In the experiment, the Fe/Cr interfaces are
 not ideally flat, but they possess steps, that lead to a random 
 growth front of the Cr spacer. One can deduce from the experimental
 data that early results, obtained from rough samples grown with
 sputtering, gave smaller strengths compared to more recent ones from
 smooth epitaxial films. Our calculation can be modified in two ways in order
 to include the effects of interface roughness. 
 First, we can average the
 amplitude of IEC over the variable thickness of the film. As is 
 shown in Refs.~\onlinecite{Bruno1,sd-mix1},
  this can be done by convoluting the coupling strength
 $J(t)$ for a certain thickness $t$ with a distribution (constant
 or Gaussian) over $t$ fluctuations. Consequently, the new coupling is
 \begin{equation}
 J_{rough}(D)= \sum_{t} P(D,t) J(t),
 \end{equation}
 where $D$ is the average thickness. Second, because of the roughness, 
 the in-plane component of the 
 wavevector is no longer a good quantum number.  
 A certain ${\bf k}_{\|}$ state can be reflected to 
 a range of other ${\bf k}_{\|}$'s states around it, and this leads to 
 a decrease of the 
 coupling, or even to complete sweep of the oscillations, if the
 period is short. Therefore, if the sample is
 not prepared to be smooth, the 2.2~ML period associated with
 the nesting vector will not be observed. This has been verified 
 experimentally~\cite{Parkin1,Unguris}. Interface roughness may also
 be responsible for the lack of the period of 4.70~\AA \ that we
 found in the (110) case.
 Apart from roughness, other disorder effects, 
 like misfit dislocations and strain~\cite{Bruno1}, 
 also can affect IEC in the same way. 
 These effects however do not play any significant role 
 in Fe/Cr multilayers because
 of the excellent match of the lattice constants (2.88~\AA \ for Cr,
 2.87~\AA \ for Fe). 
 
 The number that we give as the
 amplitude is really an upper bound of the actual value, since we are
 taking the absolute value of the complex number $|\Delta R|$.
 This is demonstrated in Fig.~\ref{f:fig5}, which
 shows $J_{ij}^{\alpha}$ vs. thickness D as calculated from Eqs.~(3) 
 and (10) for the spanning vector $\overline {N_{1}}$.
 We use $|\Delta R|$  to estimate the strength of IEC, because there is a 
 certain phase arbitrariness in the
 calculation. The main source for this is the finite thickness of
 the Fe layers, which is not taken into account in our calculation.
 It has been predicted from theory~\cite{Bruno2,Barnas}
 and verified experimentally~\cite{Okuno},
 that IEC exhibits oscillations, not only as a function
 of Cr thickness, but also as one varies the thickness of the Fe layers.
 This means of course that the antiferromagnetic peaks occur at different
 D for different Fe widths, and so one needs to use the right D in each 
 case. Moreover, for finite thickness, quantum well states are expected
 to be formed also in the magnetic layers and hence the coupling
 strength will be affected as well. Finally, we must point out once more that 
 our approximations do not hold for very thin multilayers, and
 hence the initial phase is pretty arbitrary. 
 
 Our results indicate that  the nesting vector of the ellipsoid is a favorite 
 candidate for the long period oscillations. 
 However, other candidates have been
 proposed in the literature as well. 
 Early on, it was realized that the nesting vector can
 be responsible, not only for the small, but also for the large wavelength
 oscillations. Because of the discreteness of the thickness D, a short 
 period can give rise to an effective long period as a higher harmonic.
 The phenomenon is called aliasing and one gets for the second harmonic
 a period $\lambda '=\lambda /(\lambda - 2)$~ML. For 
 $\lambda \approx 2.2$~ML the effective period is indeed large. However,
 careful measurements of the large period by 
 Pierce $et \ al.$~\cite{Unguris} 
 gave an effective period $\lambda '= 20.05$~ML, much bigger than
 the observed period which is between 10 to 12~ML. Moreover, the fact that
 the period seems to be independent of the orientation is not in favor
 of the aliasing scenario. 
 Of course from the theoretical point of view, a 
 conclusive analysis would be to calculate the
 coupling strength for the second harmonic and see 
 if it is significant.
 The nesting vector though is a difficult case, as we mentioned in
 Sec.~III. If the coupling strength for the second harmonic turns
 out to be comparable to the one of the vector $\overline {N_{1}}$, then
 the superposition of the two oscillations will lead to new features in
 the overall IEC, particularly it will raise its period and also decrease
 its amplitude.
 
 Two other candidates for the long-period coupling are the lens vectors
 $\overline {L_{1}}$ and $\overline {L_{2}}$. In a recent paper,
 Li $et \ al.$~\cite{Li} have studied the confinement of electrons 
 in the spacer using
 angle-resolved photoemission. They found substantial confinement for
 states with ${\bf k}_{\|}$ in the vicinity of the lens. 
 From this they concluded
 that the lens is the k-space origin of the long period oscillations.
 From our calculations we have found that the confinement at the lens area
 is big enough that one could detect quantum well states for this ${\bf k}_{\|}$
 (the magnitude of both $r^{+}$ and $r^{-}$ in the ferromagnetic 
 configuration is $\approx 0.5$). However, the confinement is much stronger
 at the ellipsoid as one can see in Fig.~\ref{f:fig2}(a). 
 Hence it would be useful
 to get more data for this k-space area. The confinement is one
 important factor for IEC, but there are also other factors that
 affect the amplitude of the oscillation. Most important one is the 
 spin-asymmetry of the reflection amplitude (as a difference of
 two complex numbers), and not individual magnitude by itself. The geometrical
 weight of the vector is a crucial factor as well, at least within the
 stationary phase approximation. For the lens vectors, $G_{ij}^{\alpha}$'s
 are very small. 

\section{Summary}
 
 By using an empirical TB method (including $s$, $p^{3}$, and
 $d^{5}$ orbitals) and including the spin-orbit interaction, 
 we have calculated the reflection amplitudes and 
 transmission coefficients for an Fe/Cr/Fe sandwich structure. Based
 on the spin-asymmetry of these quantities and the geometrical weight
 of the extremal vectors of the Cr Fermi surface, we have obtained the 
 coupling strengths associated with each vector. For both (001) and
 (110) growth direction, the origin of the long period oscillations
 of the exchange coupling is attributed to extremal vectors of the
 ellipsoid centered at point N. The period agrees satisfactorily with
 the observed value. The coupling strength is roughly an order of magnitude
 bigger than the experimental one, but the inclusion of interface roughness in
 the calculation can improve the agreement substantially. Although the
 attribution of the long period-coupling to the aliasing effect of the
 nesting vector faces several opposing arguments, one should try to
 calculate more carefully the coupling strength of the second harmonic of 
 this vector. Within the approximations that we have used,  
 all the other candidates for the long period (and in
 particular the lens vectors) have been found to have a small 
 to negligible contribution to the overall coupling.

\acknowledgments

This work was supported by 
the University of Illinois research board and the University of Illinois 
Materials Research Laboratory through Contract No. NSF/DMR-89-20538.

\mbox{}


\begin{figure}
\caption{(100) cross section of Cr Fermi surface.}
\label{f:fig1}
\end{figure}

\begin{figure}
\caption{Moduli of the reflection amplitudes vs. energy for (001) orientation.
 (a) ${\bf k}_{\|}=(0,0.5) \frac {2 \pi} {a}$, (b)  
${\bf k}_{\|}=(0,0.42) \frac {2 \pi} {a}$ \ (Fermi energy is aligned to be 
zero).}
\label{f:fig2}
\end{figure}

\begin{figure}
\caption{Band structure along line XN (${\bf k}_{\|}=(0,0.5)\frac {2\pi} a$
for (a) Cr, (b) majority Fe, and 
(c) minority Fe.}
\label{f:fig3}
\end{figure}

\begin{figure}
\caption{Moduli of the reflection amplitudes vs. energy for (110) orientation
and ${\bf k}_{\|}=(0.71,0)\frac {2 \pi} {a}$ \ (Fermi energy is 
aligned to be zero).}
\label{f:fig4}
\end{figure}

\begin{figure}
\caption{Coupling strength $J^{\alpha}$ \ vs. thickness D, as calculated
from a) Eq.~(3) (solid line), b) Eq.~(10) (dashed line).}
\label{f:fig5}
\end{figure}

\begin{table}
  \caption{Results for (001) orientation.}
\vspace{0.1in}
\begin{tabular}{c c d d d d} 
vector & ${\bf k}_{||}$\ [$2 \pi/a$]   & $\lambda$\, (\AA) & $G^{\alpha}$\ [meV]
  & $|\Delta R|$  &  $J^{\alpha}$ 
  (mJ/m$^2$) \\   \hline
                    & (0,0) & 5.56  & 8.48 & 1.01 & 0.68  \\  
                    & (0,0) & 9.20 & 160.12 & 9.0 10$^{-5}$ & 2.3 10$^{-3}$ \\
                    & (0,0) & 4.18 & 1.87 & 0.88 & 0.13 \\
                    & (0,0) & 33.65 & 3.40 & 6.9 10$^{-3}$ & 3.8 10$^{-3}$ \\
$\overline {V_{n}}$ & (0,0.25) & 3.12 & 146.88 & 0.40 & 37.99   \\
$\overline {L_{1}}$ & (0,0.29) & 20.85 & 1.00 & 1.7 10$^{-3}$ & 5.3 10$^{-4}$ \\
$\overline {L_{2}}$ & (0,0.29) & 17.77 & 1.39 & 0.41 & 0.18  \\
$\overline {X}$     & (0,0.42) & 11.21 & 36.99 & 0.24 & 2.81 \\
$\overline {N_{1}}$ & (0.0,0.5) & 13.77 &  24.06 & 1.03 & 7.94  \\
$\overline {N_{2}}$ & (0.5,0.5) & 10.38 & 10.40 & 0.63 & 1.03 \\
                    & (0,1.0)   & 3.42 & 6.33  & 0.65 & 0.33 \\   
 \end{tabular}
\label{table1}
\end{table}

\begin{table}
\caption{Results for (110) orientation.}
\vspace{0.1in}
\begin{tabular}{c c d d d d}  
vector & ${\bf k}_{\|}$\ [$2 \pi/a$]   & $\lambda$\, (\AA) & $G^{\alpha}$\ [meV]
  & $|\Delta R|$ &  $J^{\alpha}$ 
  (mJ/m$^2$) \\   \hline
                     & (0,0)    & 5.86  & 18.78 & 0.58 & 0.86   \\  
$\overline {L_{1}'}$ & (0,0.29) & 21.26 & 1.37 & 0.97 & 0.21  \\
$\overline {L_{2}'}$ & (0,0.29) & 17.87 & 1.57 & 1.11 & 0.28 \\
$\overline {L_{3}'}$ & (0.23,0) & 29.68 & 3.86 & 6.0 10$^{-3}$ & 3.6 10$^{-3}$\\
$\overline {X'}$     & (0,0.44) & 9.42  & 6.01 & 0.93 & 0.89 \\
$\overline {N_{1}'}$ & (0.71,0) & 16.46 & 38.45 & 1.51 & 4.65 \\
$\overline {N_{2}'}$ & (0.35,0.50) & 12.12 & 15.41 & 0.24 & 1.17 \\
$\overline {N_{3}'}$ & (0.71,1.0)   & 10.04 & 7.67 & 0.25 & 0.15 \\ 
                     & (0.3,0.0) & 9.73 & 12.93 & 0.07 & 0.30 \\
$\overline {S'}$     & (0,1.0)   & 4.70 & 43.31 & 0.59 & 2.02 \\ 
\end{tabular}
\label{table2}
\end{table}

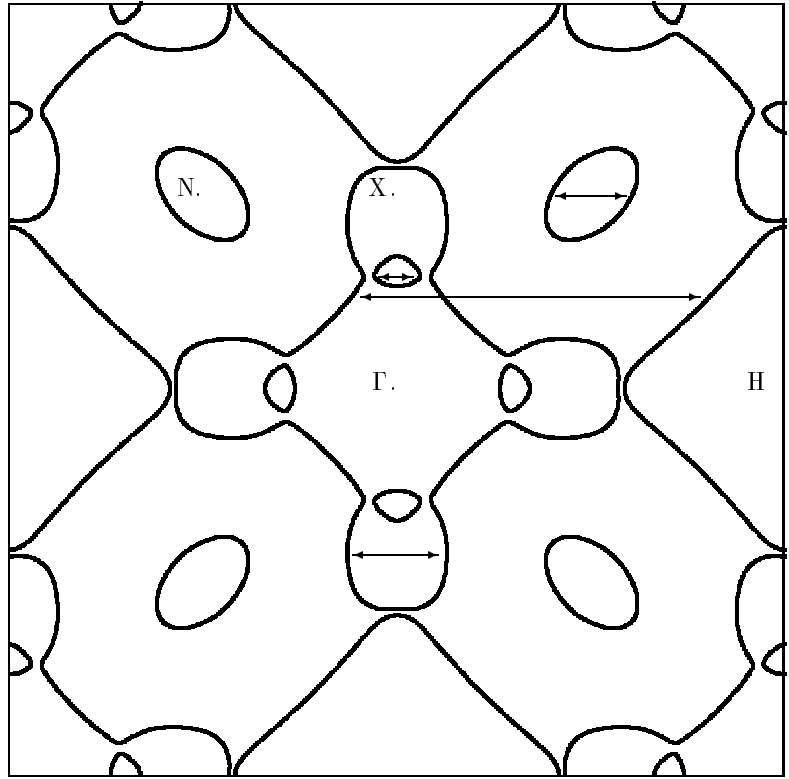
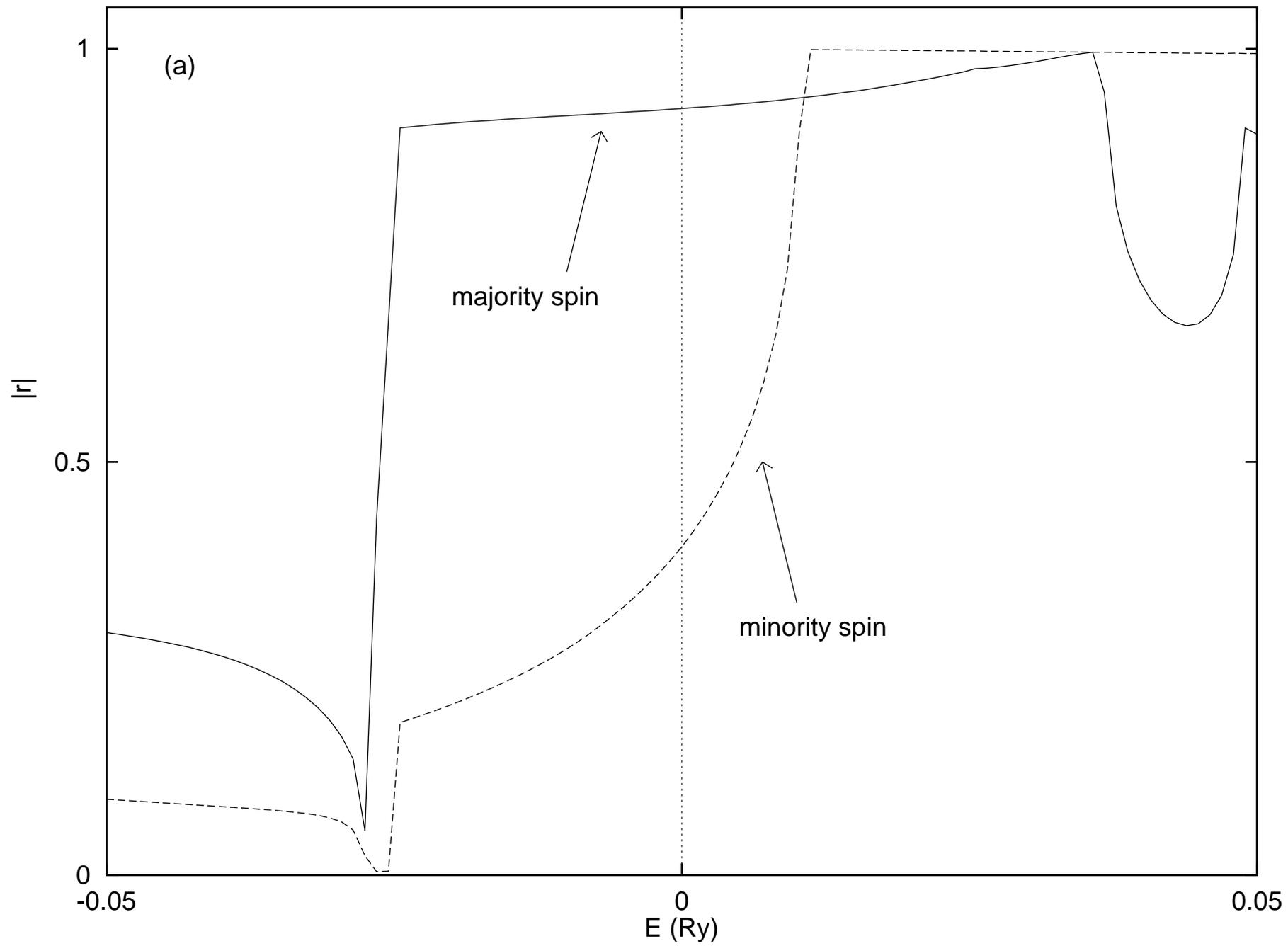
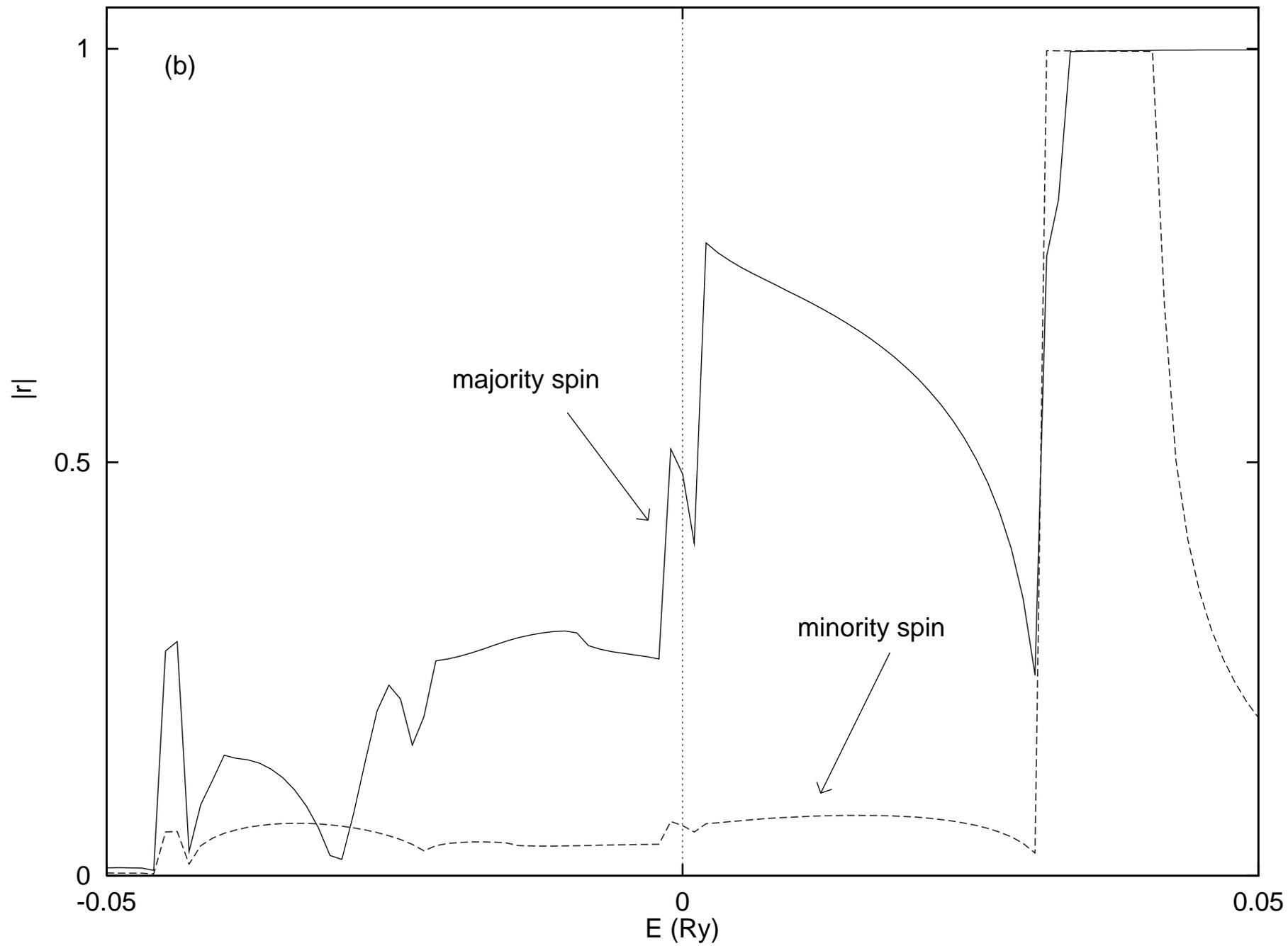
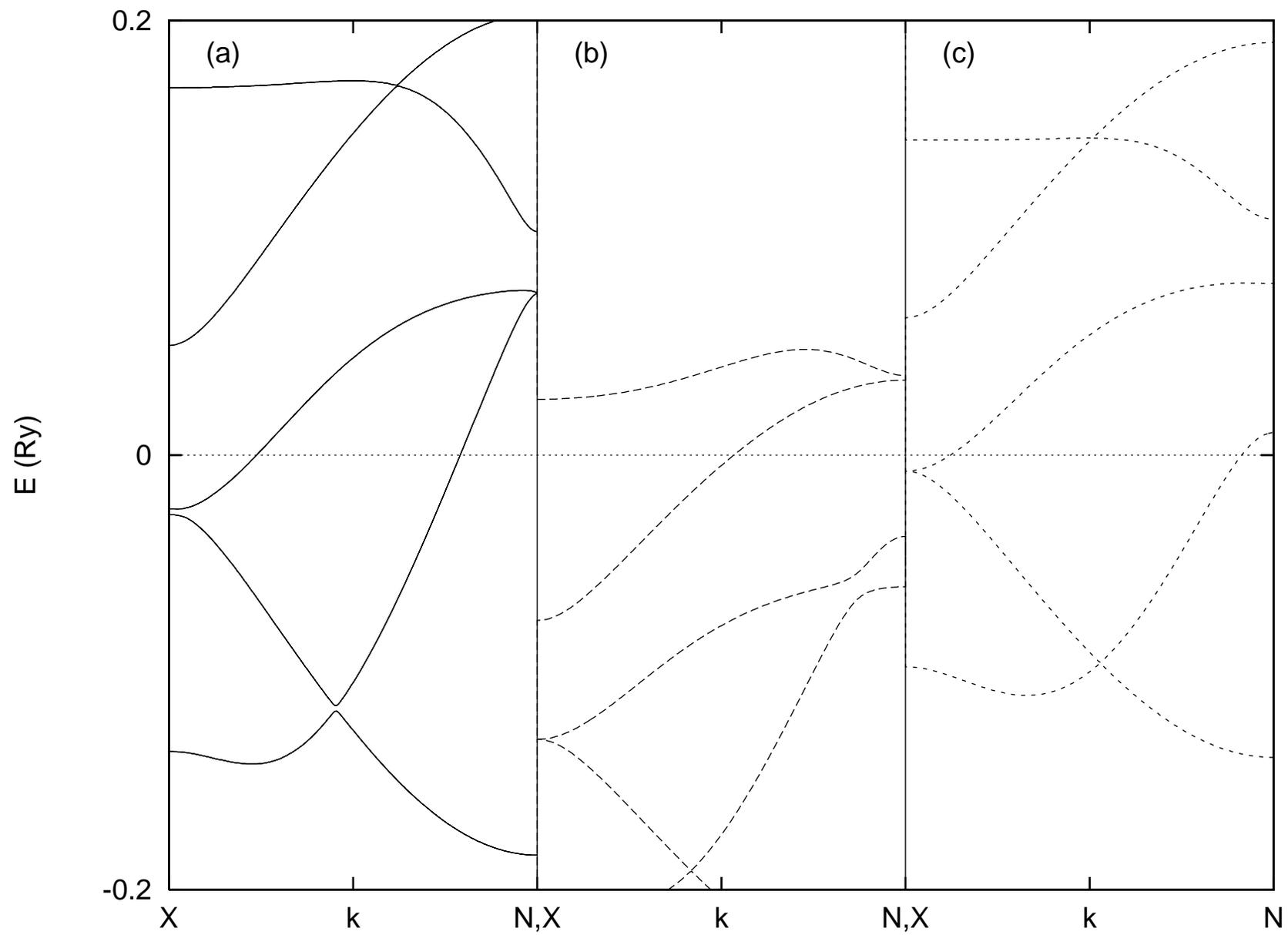
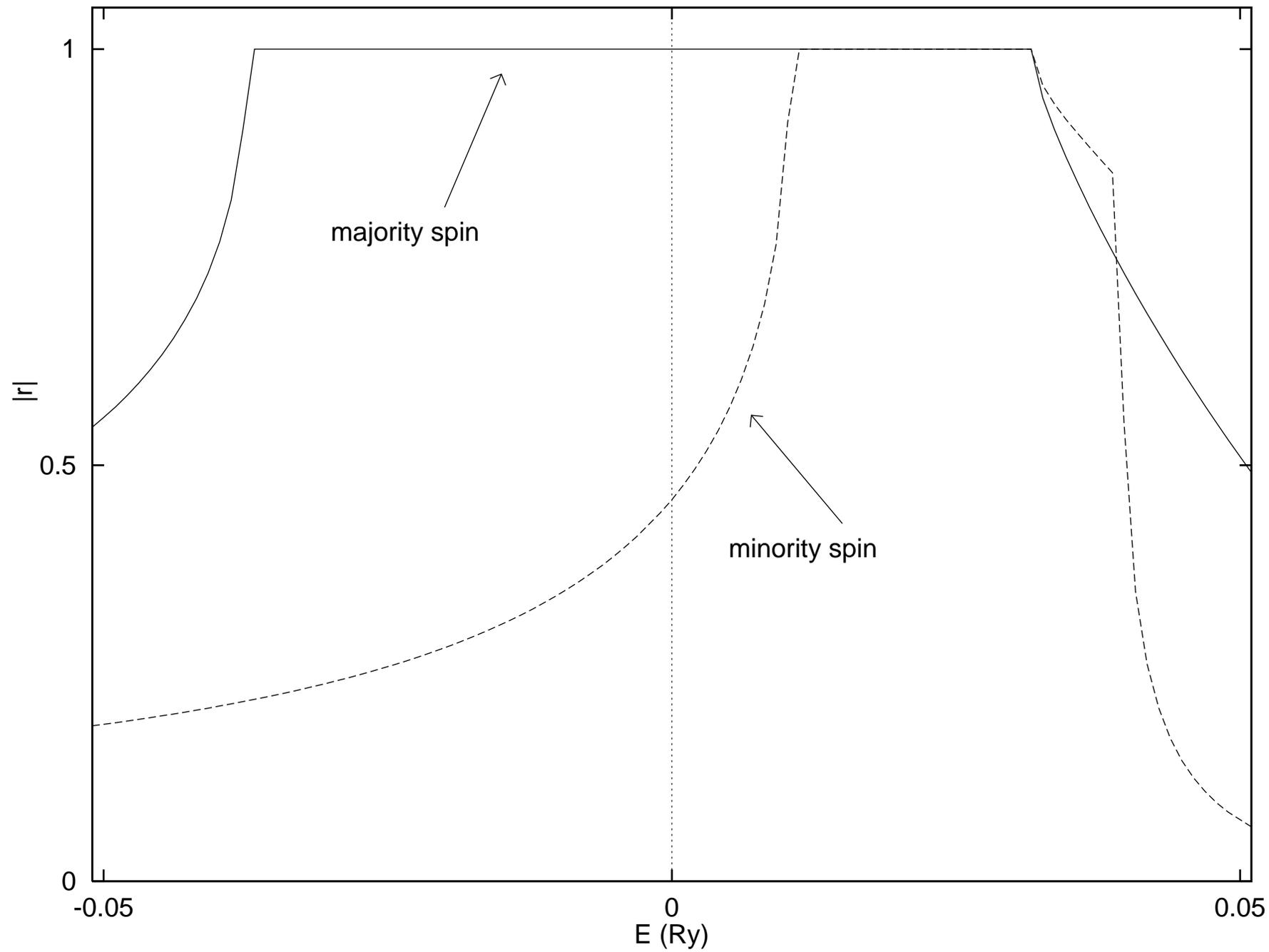
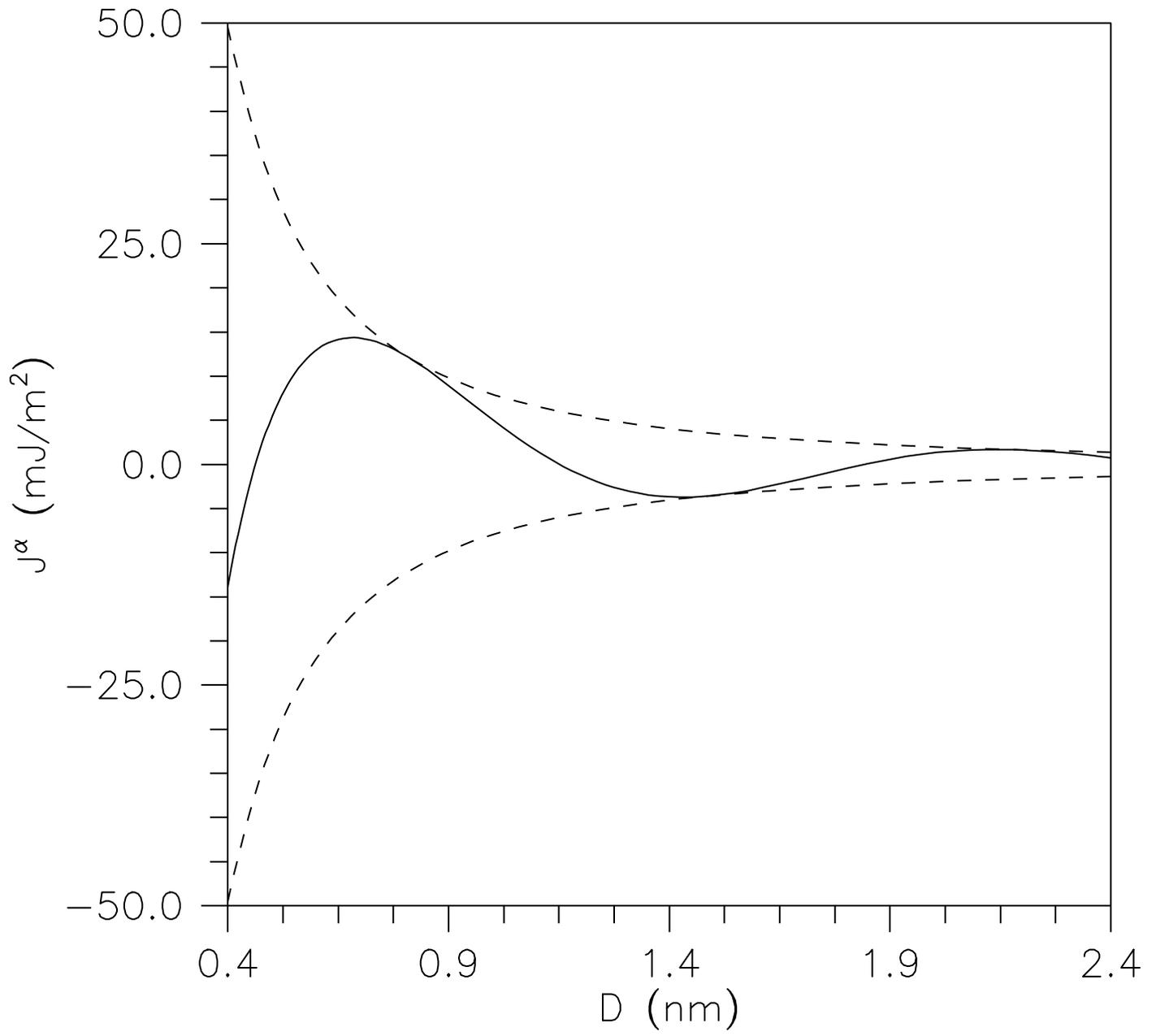
\end{document}